\documentclass[12pt]{article}

\usepackage{amsmath}
\usepackage{amssymb}

\newtheorem{theorem}{Theorem}[section]

\newtheorem{corollary}{Corollary}[section]

\newtheorem{definition}{Definition}[section]
\numberwithin{equation}{section}

\begin{document}

\begin{center}
 Paragrassmann Algebras as
Quantum Spaces 
\\
Part II:  Toeplitz Operators
 \vskip 0.5cm
 Stephen Bruce Sontz
  \\
 Centro de Investigaci\'on en Matem\'aticas, A.C.
 \\
 CIMAT
  \\
 Guanajuato, Mexico
  \\
 email: sontz@cimat.mx
 \end{center}

\begin{abstract}
\noindent
This paper continues the study of paragrassmann algebras 
begun in Part I with the definition and analysis of
Toeplitz operators in the associated holomorphic Segal-Bargmann space.
These are defined in the usual way as multiplication by a symbol
followed by the projection defined by the reproducing kernel. 
 These are non-trivial examples of spaces with
Toeplitz operators whose symbols are not functions and which themselves are not spaces of functions.

\end{abstract}



\section{Introduction}

The paper is organized in three sections. Here in the first
we briefly survey some relevant material from Part I \cite{part1}. 
The intention is that the notation in this paper be identical to that in \cite{part1}.
The material in this section will be used repeatedly without explicit reference.
See \cite{part1} for further details.
Then in the second section we present our results in full.
And in the last section we comment about possibilities for future research.
For background information on this area of research in physics and mathematics,
see \cite{csq} and references given there.
Throughout this paper we let  $l \ge 2$ be an integer and $q \in \mathbb{C} \setminus \{ 0 \}$.

\begin{definition}
The {\rm paragrassmann algebra} $PG_{l,q}(\theta, \overline{\theta})$
with {\rm paragrassmann variables} $\theta$ and $\overline{\theta}$ is defined to be
\begin{equation*}
PG_{l,q} =
PG_{l,q}(\theta, \overline{\theta})  := \mathbb{C}  \{ \theta, \overline{\theta} \} / J.
\end{equation*}
\end{definition}
Here $J$ is the two sided ideal generated by $\theta^l, \overline{\theta}{}^l, \theta \overline{\theta}
- q \overline{\theta} \theta $
and $\mathbb{C}  \{ \theta, \overline{\theta} \} $ is
the free algebra over the field of complex numbers $\mathbb{C}$ generated by
the set $\{ \theta, \overline{\theta} \}$ of two elements.
We say that $\theta$ is a \textit{holomorphic} variable while 
$\overline{\theta}$ is an \textit{anti-holomorphic} variable.

We let $I_l := \{ 0, 1, \dots , l-1 \}$ be an index set.
If we use a variable, such as $i$, without specifying its domain, it is understood
that $i \in I_l$.

We have $\dim_{\mathbb{C}} PG_{l,q}(\theta, \overline{\theta}) = l^2$, which is left to the
reader as an exercise.
We will be using the \textit{anti-Wick} basis
$ AW = \{ \theta^i \overline{\theta}{}^j  \, | \, i,j \in I_l \}$ of $PG_{l,q}(\theta, \overline{\theta})$.

The \textit{Segal-Bargmann space} (or \textit{holomorphic space}) is defined to be
\begin{equation*}
\mathcal{B}_H = \mathcal{B}_H(\theta) := \mathrm{span}_{\mathbb{C}} \, \{ \theta^i \, | \, i \in I_l \}
\subset PG_{l,q}(\theta, \overline{\theta}).
\end{equation*}
This is a commutative sub-algebra which is not isomorphic as an algebra
to an algebra of complex valued functions, since $\theta \ne 0$ is nilpotent.
On the other hand, the \textit{anti-Segal-Bargmann space} (or \textit{anti-holomorphic space}) is defined to be
\begin{equation*}
\mathcal{B}_{AH} = \mathcal{B}_{AH}(\overline{\theta}) :=
\mathrm{span}_{\mathbb{C}} \, \{ \overline{\theta}{}^i \, | \, i \in I_l \} \subset PG_{l,q}(\theta, \overline{\theta}).
\end{equation*}

We define a conjugation in $PG_{l,q}(\theta, \overline{\theta})$ by putting 
$( \theta^i \overline{\theta}{}^j)^*:=  \theta^j \overline{\theta}{}^i$
for the basis elements in $AW$ and then extending \textit{anti-linearly} to $PG_{l,q}(\theta, \overline{\theta})$.
This conjugation satisfies the $*$-algebra condition 
(which says that $(fg)^* = g^* f^*$ for all $f,g \in PG_{l,q}(\theta, \overline{\theta}) \,$) if and only if
$q \in \mathbb{R} \setminus\{ 0 \}$. 
However, the $*$-algebra condition is not used anywhere in this paper, which is why
we consider the more general situation $q \in \mathbb{C} \setminus\{ 0 \}$. 

We define a Berezin type integral that maps $PG_{l,q}$ to $\mathbb{C}$ by setting
$$
\int  \!\!\!  \int d \theta   \,\, 
\theta^i \overline{\theta}{}^j
 \,\,  d \overline{\theta} := \delta_{i,l-1} \delta_{j,l-1}
$$
for basis elements in $AW$ and then extending \textit{linearly} to $PG_{l,q}$.
Here $\delta_{n,m}$ is the Kronecker delta for the integers $n,m$.

Take $f = f(\theta, \overline{\theta})$ and $ g = g(\theta, \overline{\theta})$ in $PG_{l,q}$.
Let $w_n$ for $n  \in I_l$ be a finite sequence of strictly positive real numbers.
These can be thought of as `weights' if one likes.
We define
\begin{equation*}
\langle f , g \rangle_w := \sum_{m \in I_l} w_{l-1-m} \int \!\!\! 
\int d \theta   \,\, \theta^m
              : f(\theta, \overline{\theta})^* : \, : g(\theta, \overline{\theta}) : 
\overline{\theta}{}^m \,\,  d \overline{\theta},
\end{equation*}
where the \textit{anti-Wick product} $\, : \cdot : \, : \cdot : \,$
is defined as the $\mathbb{C}$-bilinear extension to $PG_{l,q} \times PG_{l,q}$  of 
\begin{equation*}
: \theta^a \overline{\theta}{}^b : \, : \theta^c \overline{\theta}{}^d : \,\,\, \equiv \,\,\, \theta^{a+c} \overline{\theta}{}^{b+d} 
\end{equation*}
for any pair of elements of $AW$.
The anti-Wick product maps $PG_{l,q} \times PG_{l,q}$ to $PG_{l,q}$.
An important consequence of this definition is
$$
 \langle \theta^a \overline{\theta}{}^b ,  \theta^c \overline{\theta}{}^d \rangle_w 
 = \delta_{a+d, b+c} \, \chi_l (a+d) \, w_{a+d},                             
$$
where $\chi_l$ is the characteristic function of $I_l$, that is for every integer $k$ we put $\chi_l(k) = 1$ if $k \in I_l$
and $\chi_l(k) = 0$ if $k \notin I_l$.
Also we put $w_n := 0$ if $n \ge l$.
Another consequence of this definition is that there is always an element $f \ne 0$ such that $\langle f , f \rangle_w = 0$
and another element $g$ such that $\langle g , g \rangle_w < 0$.
However, the inner product $\langle \cdot , \cdot \rangle_w $ restricted to the
Segal-Bargmann space $\mathcal{B}_H(\theta) $ is a positive definite inner product with
$\langle \theta^j , \theta^k \rangle_w = \delta_{j,k} w_j$.
So the weight $w_j$ is associated to the monomial $\theta^j $.
So it turns out that 
$$
\phi_j (\theta) := w_j^{-1/2} \, \theta^j 
$$ 
is an orthonormal basis of $\mathcal{B}_H(\theta) $,
where we take the positive square root of $w_j > 0$.
We define $ || f ||^2_w := \langle f , f \rangle_w$ for any $f \in PG_{l,q}$.

The reproducing kernel for the Segal-Bargmann space $\mathcal{B}_H(\theta) $ exists and is unique.
It is given by
$$
        K(\theta, \eta) = \sum_j \dfrac{1}{w_j} \overline{\theta}{}^j \otimes \eta^j,
$$
where $\eta$ is another paragrassmann variable.
Specifically, for every element $f(x) \in \mathbb{C}[x]$, the polynomial ring in $x$, we have the \textit{reproducing formula}
\begin{equation*}
    f(\theta) = \langle K(\theta, \eta) , f(\eta) \rangle_w,
\end{equation*}
where the inner product is taken with respect to the variables $\eta, \overline{\eta}$.
If 
$$
f(x) = \sum_{j=0}^N \alpha_j x^j  \in \mathbb{C}[x],
$$
 where $\alpha_j \in \mathbb{C}$, then
$f(\theta) := \sum_{j=0}^N \alpha_j \theta^j$ defines a functional calculus that plays the role here played by
`evaluation at a point' in the classical theory of reproducing kernel Hilbert spaces of functions.
Similarly, $f(\eta) := \sum_{j=0}^N \alpha_j \eta^j$.
Of course, in this case we have that 
$$
f(\theta) = \sum_{j=0}^{\min (N,l-1)} \alpha_j \theta^j.
$$

Also, we have a canonical linear isomorphism
$\delta_{\theta \to \eta} : \mathcal{B}_H(\theta) \to
 \mathcal{B}_H(\eta)$ induced by $\theta^i \mapsto \eta^i$ for all $i \in I_l$.

\section{Toeplitz operators}

The reproducing kernel on the Segal-Bargmann space
allows us to define Toeplitz operators in more or less the usual way.
The only subtle point is that multiplication operators can be defined as acting on the left or on the right.
The basic idea is to take an arbitrary element $g \in PG_{l,q}(\eta, \overline{\eta})$  and
 define $\tilde{T_g} : \mathcal{B}_H(\eta) \to \mathcal{B}_H(\theta)$ by a formula such as                                                                             
\begin{equation}
\label{f-left-first-def-toeplitz}
\tilde{T}_g f (\theta) : = \langle K(\theta, \eta) \, , \,  f(\eta) \, g(\eta, \overline{\eta}) \rangle_w,
\end{equation} 
where $f \in \mathcal{B}_H(\eta) $.
Or, if we wish, we could define it by
\begin{equation}
\label{f-right-first-def-toeplitz}
T^{\#}_g f (\theta) : = \langle K(\theta, \eta) \, , \, g(\eta, \overline{\eta}) \, f(\eta) \rangle_w,
\end{equation}
where $f \in \mathcal{B}_H(\eta) $.
In fact, we will take (\ref{f-left-first-def-toeplitz}) to be our preliminary definition of a Toeplitz operator $\tilde{T}_g$
with \textit{symbol} $g(\eta, \overline{\eta})$.
The definition (\ref{f-right-first-def-toeplitz}) gives a similar theory (though with some extra factors of $q$), so we do not
develop it here.

Of course, this is a quantization scheme in a generalized sense since it sends `functions' $g$ to
operators $\tilde{T_g}$. 
We recall here the famous criterion that ``quantization is operators instead of functions.''
And even though this operator does not map a Hilbert space to itself, we can precompose (or
postcompose) $\tilde{T_g}$ with the canonical isomorphism $\delta_{\theta \to \eta}$ to get an operator from
 either of these two Hilbert spaces to itself.
 Simply as a convention 
 we take $T_g := \tilde{T}_g \circ \delta_{\theta \to \eta} : \mathcal{B}_H(\theta) \to
 \mathcal{B}_H(\theta)$ as our definition of the
 \textit{Toeplitz operator} $T_g$ with \textit{symbol} $g = g(\eta, \overline{\eta}) \in PG_{l,q}(\eta, \overline{\eta})$.
 Since we have a functional calculus of $\eta,\overline{\eta} \in PG_{l,q}(\eta, \overline{\eta})$
 using the non-commuting polynomials
 $g \in \mathbb{C}  \{ \eta, \overline{\eta} \} $, we can also use $ \mathbb{C}  \{ \eta, \overline{\eta} \} $
 as the space of symbols here instead of using $PG_{l,q}(\eta, \overline{\eta})$.
  
 We remark that these Toeplitz operators (which form a quantization) should not be confused
 with the operators $A_f$ introduced in \cite{csq} nor with that quantization.
 Those operators $A_f$ are called the \textit{coherent state quantization} and rely on the resolution of the identity
provided by the coherent states.
Here we are using the reproducing kernel to project after multiplying by a symbol, and this
is the standard definition of a Toeplitz operator in analysis.
(See \cite{hall} or \cite{rama-pr}.)
These are conceptually quite different constructs. 
For the coherent state quantization one needs a measure even to be able 
to write down the integral formula for the resolution of the identity and then the
corresponding integral formula for the definition of $A_f$.
Or, as is the case in \cite{csq} and in this paper, one needs some reasonable generalization
of an integral instead of a measure \textit{per se}.
For the Toeplitz quantization one only needs a reproducing kernel, whose basic
reproducing property requires an inner product. 
And that inner product does not necessarily arise from a measure. 
However, two conceptually distinct constructions could turn out to be equivalent, at least in this
particular case. We will return to this question.

Before studying the Toeplitz operators themselves, let us consider the operator associated with the reproducing
kernel $K$. We define this operator
$P_K : PG_{l,q}(\theta, \overline{\theta}) \to PG_{l,q} (\theta, \overline{\theta})$ by
\begin{equation}
\label{define-PK}
P_K F (\theta) := \langle K(\theta, \eta) , F(\eta, \overline{\eta}) \rangle_w
\end{equation}
for all $ F(\theta, \overline{\theta}) \in PG_{l,q}(\theta, \overline{\theta})$.
It is clear that $P_K F (\theta)$ lies in the Segal-Bargmann space $\mathcal{B}_H (\theta)$ and
that the range of $P_K$ is $\mathcal{B}_H (\theta)$ since $P_K$ restricted to $\mathcal{B}_H (\theta)$
is the identity map.
The classical theory suggests that $P_K$ is an orthogonal projection.
And this is so.

\begin{theorem}
The linear map $P_K$ is an orthogonal projection, that is, it satisfies
$P_K = P_K^2 = P_K^*$.
\end{theorem}
\textbf{Remark}: The adjoint $P_K^*$ is defined with respect to the nondegenerate sesquilinear form
$\langle \cdot , \cdot \rangle_w$ on $PG_{l,q}(\theta, \overline{\theta}) $.
(See Theorem 8.1 in \cite{part1}.)
Therefore adjoints exist and are unique with respect to that form.
Even though 
$P_K^2 = P_K$ follows from the comments above, we prove it explicitly anyway.

\vskip 0.2cm \noindent
\textbf{Proof}:
We will use the notation $F_{ab} := \theta^a \overline{\theta}{}^b$ for the elements in the basis $AW$.
Acting with $P_K$ on this basis we obtain
\begin{gather*}
P_K F_{ab} (\theta) =  \langle K(\theta, \eta) , F_{ab} (\eta, \overline{\eta}) \rangle_w
=  \langle K(\theta, \eta) , \eta^a \overline{\eta}{}^b \rangle_w
\\
= \sum_k \dfrac{1}{w_k} \langle  \eta^k , \eta^a \overline{\eta}{}^b \rangle_w \, \theta^k
= \sum_k \dfrac{1}{w_k} \delta_{k+b,a} \chi_l(a) \chi_l(k) w_a \, \theta^k = \dfrac{w_a}{w_{a-b}} \chi_l(a-b) \theta^{a-b}.
\end{gather*}
Applying $P_K$ twice we have that
\begin{gather*}
P_K^2 F_{ab} (\theta) = P_K \left( \dfrac{w_a}{w_{a-b}} \chi_l(a-b) \theta^{a-b} \right)
 =  \dfrac{w_a}{w_{a-b}} \chi_l(a-b) P_K \left(\theta^{a-b} \overline{\theta}{}^0 \right)
 \\
 =  \dfrac{w_a}{w_{a-b}} \chi_l(a-b) \dfrac{w_{a-b}}{w_{a-b}} \chi_l(a-b) \theta^{a-b}
 =  \dfrac{w_a}{w_{a-b}} \chi_l(a-b) \theta^{a-b}
 = P_K F_{ab} (\theta),
\end{gather*}
which implies that $P_K^2 = P_K$.

Next here are the matrix elements for $P_K$:
\begin{gather*}
\langle F_{ab}, P_K F_{cd} \rangle_w =
\langle \theta^a \overline{\theta}{}^b, \dfrac{w_c}{w_{c-d}} \chi_l(c-d) \theta^{c-d}
\rangle_w 
\\ 
= \dfrac{w_c}{w_{c-d}}  \chi_l(c-d)  \delta_{a, b+c-d} \chi_l(a) w_a
= \dfrac{w_a w_c}{w_{c-d}} \chi_l(c-d) \delta_{a-b,c-d}.
\end{gather*}

Now we calculate the matrix elements for $P_K^*$ obtaining
\begin{gather*}
\langle F_{ab}, P_K^* F_{cd} \rangle_w = \langle P_K F_{ab},  F_{cd} \rangle_w =
\langle \dfrac{w_a}{w_{a-b}} \chi_l(a-b) \theta^{a-b}, \theta^c \overline{\theta}{}^d \rangle_w 
\\
= \dfrac{w_a}{w_{a-b}}\chi_l(a-b) \delta_{a-b+d,c} \chi_l(c) w_c
= \dfrac{w_a w_c}{w_{a-b}} \chi_l(a-b) \delta_{a-b,c-d}.
\end{gather*}
Since these matrix entries for  $P_K^*$ are equal to those for  $P_K$,
we conclude that $P_K^* = P_K$.
$\quad \blacksquare$

\vskip 0.4cm \noindent
For any $g \in PG_{l,q}$ we define the linear map 
$M_g : PG_{l,q}(\theta, \overline{\theta}) \to PG_{l,q}(\theta, \overline{\theta})$
to be multiplication by $g$ on the right, that is
$$
        M_g F := F g
$$
for all $F \in PG_{l,q}(\theta, \overline{\theta})$.
Then we have the following result.
\begin{theorem}
Say $f_1, f_2 \in \mathcal{B}_H (\theta)$. Then we have
$$
\langle f_1, T_g f_2\rangle_w = \langle f_1, M_g f_2 \rangle_w.
$$
\end{theorem}
\textbf{Proof}:
We first note that $T_g = P_K M_g$, that is, the Toeplitz operator is
right multiplication by $g$ followed by the projection associated to $K$.
Then we calculate
\begin{gather*}
\langle f_1, T_g f_2\rangle_w = \langle f_1, P_K M_g f_2\rangle_w
= \langle P_K^* f_1,  M_g f_2\rangle_w
\\
 = \langle P_K f_1,  M_g f_2\rangle_w
 = \langle  f_1,  M_g f_2\rangle_w
\end{gather*}
as claimed.
$\quad \blacksquare$

\vskip 0.4cm
Now we note that the correspondence $g \mapsto T_g$ gives us a mapping
$$
T : PG_{l,q}(\eta, \overline{\eta}) \to \mathcal{L} ( \mathcal{B}_H(\theta) ),
$$
where $\mathcal{L} ( \mathcal{B}_H(\theta) ) $ is the $*$-algebra of all linear endomorphisms
of the Hilbert space $\mathcal{B}_H(\theta)$.
It is easily verified that $T$ is linear and that $T_1 = I_{\mathcal{B}_H(\theta)}$, the identity.
We note that the dimensions of the domain and codomain of $T$ are equal, since
$\dim PG_{l,q}(\eta, \overline{\eta}) = l^2$ while $\dim  \mathcal{L} ( \mathcal{B}_H(\theta) ) =
( \dim \mathcal{B}_H(\theta) )^2 = l^2$.
One would like to know whether $T$ is an isomorphism of vector spaces and how it relates
to the algebra structure on its domain and codomain.
Of course, the non-commutativity of $ PG_{l,q}(\eta, \overline{\eta})$ is completely determined by the
non-commutativity of the two generators $\eta, \overline{\eta}$, while the non-commutativity of
$\mathcal{L} ( \mathcal{B}_H(\theta) )$
($\cong l \times l$ complex matrices) is not so simply described in general. 
It seems that
these two algebras in general are not isomorphic as algebras,
though (as is well known) they are for $q= -1$ and $l=2$.
But we do have:

\begin{theorem}
\label{T-is-isomorphism}
The linear map
$
T : PG_{l,q}(\eta, \overline{\eta}) \to \mathcal{L} ( \mathcal{B}_H(\theta) )
$
is a vector space isomorphism.
\end{theorem}
\textbf{Proof}:
It suffices to show that the kernel of $\tilde{T}$ is zero.
So we take $g \in \ker \tilde{T}$, which means that  $\tilde{T}_g = 0$.
In particular, this implies that $\tilde{T}_g f_d = 0$ for all $d \in I_l$, where $f_d = \theta^d$.
Now, writing $g = \sum_{i j} g_{i j} \eta^i \overline{\eta}{}^j$ with $g_{i j} \in \mathbb{C}$ we calculate
\begin{gather*}
\tilde{T}_g f_d (\theta)  = \langle K(\theta, \eta) \, , \,  f_d(\eta) \, g(\eta, \overline{\eta}) \rangle_w
=   \sum_c  \dfrac{1}{w_c}  \sum_{i j} g_{i j} \langle \overline{\theta}{}^c \otimes \eta^c
 \, , \,  \eta^d \, \eta^i \overline{\eta}{}^j \rangle_w
 \\
=  \sum_c  \dfrac{1}{w_c}  \sum_{i j} g_{i j} \langle \eta^c \, , \,  \eta^d \, \eta^i \overline{\eta}{}^j \rangle_w \, \theta^c
=  \sum_c  \dfrac{1}{w_c}  \sum_{i j} g_{i j} 
\langle \eta^c \, \overline{\eta}{}^d , \, \eta^i \overline{\eta}{}^j \rangle_w \, \theta^c.
\end{gather*}
So, $\tilde{T}_g f_d = 0$
for all $d \in I_l$ implies
$
\sum_{i j} g_{i j} 
\langle \eta^c \, \overline{\eta}{}^d , \, \eta^i \overline{\eta}{}^j \rangle_w =0
$
for all $c, d \in I_l$.
These are $l^2$ homogeneous linear equations in the $l^2$ coefficients $g_{i j} $.
As we have already seen (Theorem 8.1 in \cite{part1}) the $l^2 \times l^2$ matrix
$\langle \eta^c \, \overline{\eta}{}^d , \, \eta^i \overline{\eta}{}^j \rangle_w$ is invertible and so
we must have $ g_{i j} =0$ for all $i, j \in I_l$.
Hence $g=0$.
$\quad \blacksquare$

\vskip 0.2cm \noindent
So for any non-zero $g$ in $PG_{l,q}(\eta, \overline{\eta}) $ the associated Toeplitz
operator $T_g$ is non-zero and hence its operator norm is strictly positive, $|| T_g ||_{op} >0$.
But $\langle g , g \rangle_w$ can be zero or negative. 
So there is no way to bound the operator norm above by a multiple of  $\langle g , g \rangle_w$,
nor even by  a multiple of  $|\langle g , g \rangle_w|$.
This contrasts with the situation in the classical case. (See \cite{hall}.)

Here is another useful general result about these Toeplitz operators.
\begin{theorem}
The matrix of $T_{\eta^i \overline{\eta}{}^j }$ with respect to the {\em ordered}, orthogonal basis 
$$
\big\{ \, \theta^a ~|~ a \in I_l \, \big\} 
$$
of $\mathcal{B}_H(\theta)$
(where the usual order for the integers in $I_l$ induces the order in the basis)
 is a matrix whose columns contain either all zeros or exactly one non-zero entry.
 Column $a$ of this matrix has a non-zero entry in row $a+i-j$ if and only if $a+i \in I_l$
 and $a+i-j \in I_l$.
 In this case the non-zero entry is  $w_{i+a} / w_{i+a-j} $.
 Otherwise, column $a$ contains only zeros.
\end{theorem}
\textbf{Proof}: We will examine first the image under
$\tilde{T}$ of $\eta^i \overline{\eta}{}^j \in PG_{l,q}(\eta, \overline{\eta}) $.
So, taking  $f_a(\eta) = \eta^a \in \mathcal{B}_H(\eta)$ with $a \in I_l$ we obtain
\begin{eqnarray}
 && (\tilde{T}_{\eta^i \overline{\eta}{}^j } f_a)    (\theta) =
    \langle K(\theta, \eta) \, , \, f_a(\eta) \, \eta^i \overline{\eta}{}^j    \rangle_w
      = \langle \sum_{k} \dfrac{1}{w_k} \overline{\theta}{}^k \otimes \eta^k \, , \, 
   \eta^a \eta^i \overline{\eta}{}^j  \rangle_w \nonumber
   \\ 
   &=&
    \sum_{k} \dfrac{1}{w_k} \theta^k \otimes
   \langle \eta^k \, , \, 
   \eta^{i+a} \overline{\eta}{}^j   \,  \rangle_w
    = \sum_{k} \dfrac{1}{w_k} \delta_{j+k, i+a} \chi_l(j+k) \chi_l(k) w_{j+k} \, \theta^k  \nonumber
    \\
    \label{useful}
   &=&
     \dfrac{w_{i+a}}{w_{i+a-j}} \chi_l(i+a) \chi_l(i+a-j) \, \theta^{i+a-j} .
\end{eqnarray}
Strictly speaking, we must define $\theta^k$ and $w_k$ for $k \notin I_l$.
But we can give these expressions arbitrary definitions, since the $\chi_l$ factors
give zero in this case.
Of course, we already have $\theta^k =0$ for $k \ge l$.
We also define $\theta^k =0$ for $k < 0$ for convenience. 

So $ T_{\eta^i \overline{\eta}{}^j }$ sends the basis element $\theta^a$ to a multiple
of the element $\theta^{i+a-j} $, which is a basis element exactly when $i+a-j \in I_l$.
(Otherwise, $\theta^{i+a-j} = 0$.)
This multiple is non-zero exactly when we also have $a+i \in I_l$.
And by linear algebra the coefficients of the expansion of the image of $\theta^a$ in the basis
$\theta^b$, $b \in I_l$, are the entries in column $a$ of the matrix associated to $ T_{\eta^i \overline{\eta}{}^j }$.
$\quad \blacksquare$

\vskip 0.4cm \noindent
\textbf{Remark}: If instead we had used the orthonormal basis $\phi_a(\theta)$,  $a \in I_l$, then the associated matrix 
would have non-zero entries in exactly the same entries, only now the non-zero entry in column $a$
would become $w_{a+i}/ ( w_{a}w_{a+i-j} )^{1/2}$ provided that
both $a+i \in I_l$ and $a+i-j \in I_l$ hold.
\vskip 0.3cm

We now derive some consequences of (\ref{useful}).
In particular, when $i=j=0$ we have that $ T_{\eta^i \overline{\eta}{}^j }= T_1$ sends $\theta^a$ to itself for all $a \in I_l$.
And this agrees with the fact, noted earlier, that $T_1 = I$, the identity.
For the more general case $i=j$ we have that $ T_{\eta^i \overline{\eta}{}^i }$ sends
$\theta^a$ to $ (w_{i+a} / w_{a}) \theta^a$ for $a = 0, 1, \dots , l-1-i$, while it sends $\theta^a$ to zero for $a \ge l-i$.
So the matrix of $ T_{\eta^i \overline{\eta}{}^i }$  in the above basis is diagonal with non-negative eigenvalues 
and rank equal to $l-i$.

By taking $j=0$ in (\ref{useful}), we have that
$$
       T_{\eta^i} : \theta^a \mapsto \chi_l(a+i) \theta^{a+i} = \theta^{a+i},
$$ 
which we can write as $T_{\eta^i} = M_{\theta^i}$, the operation
of multiplying by $\theta^i$.
(Since $\mathcal{B}_{H} (\theta)$ is commutative, we do not have to distinguish between right and
left multiplication.)
In particular, $T_{\eta^i} = (T_{\eta})^i$ for all $i \in I_l$.
Of course, this is as it must be since $T_{\eta} $ is really multiplication by $\theta$ (which
maps $\mathcal{B}_{H} (\theta)$ to itself) followed by the projection induced by the reproducing
kernel, which acts as the identity on $\mathcal{B}_{H} (\theta)$.
Clearly, $T_{\eta} $ is not invertible, and its kernel is given by $\ker T_{\eta} = \mathbb{C} \, \theta^{l-1}$.
It easily follows that $ (T_{\eta})^i \ne 0$ for $0 \le i < l$ and that $ (T_{\eta})^l = 0$.
So $T_{\eta}$ is nilpotent of order $l$.
Clearly, the family of operators $T_{\eta^i} $ for $i \in I_l$ is commutative, since each one is a power of
a fixed operator, namely $T_{\eta} $.

Coming back to the problem of estimating the operator norm of a Toeplitz operator, we note that the above shows that
$  T_{\eta} ( \phi_a )= (  w_{a+1}/w_a)^{1/2} \phi_{a+1}$. This implies
$ || T_{\eta}  ||_{op}^2 \ge \max_a (  w_{a+1}/w_{a} ) $, while $ || \eta ||_w^2 = w_1 $.
(Here $w_l =0$.)
So even for a holomorphic symbol, estimating the operator norm depends on information about all
the weights $w_n$ which are $l$ independent parameters.

The case $i=0$ in (\ref{useful}) is a bit different.
(Note that we have already discussed the case $i=j=0$.)
In this case we have
$$
       T_{\overline{\eta}{}^j} : \theta^a \mapsto \dfrac{w_a}{w_{a-j}} \theta^{a-j}.
$$
In particular 
$$
       T_{\overline{\eta}} : \theta^a \mapsto \dfrac{w_a}{w_{a-1}} \theta^{a-1},
$$
which can be viewed as a weighted shift operator or as a generalized derivative operator.
Taking $a=0$, we have that $T_{\overline{\eta}} : 1 \mapsto 0$, and so $T_{\overline{\eta}}$
is not invertible.
We have that $\ker T_{\overline{\eta}} = \mathbb{C} \, 1$.
We note that $ T_{\overline{\eta}{}^j} =  (T_{\overline{\eta}})^j$ follows immediately.
As above, we easily see that  $T_{\overline{\eta}}$ is nilpotent of order $l$.
We will use the notation $\partial_w := T_{\overline{\eta}}$ to indicate that
this is a type of derivative.
Note that the parameters $w_a$ arise from the definition of the
sesquilinear form which are, in general, independent from the other parameter $q$.
We also note that the family of operators $T_{\overline{\eta}^i} $ for $i \in I_l$ is commutative, since again each one is a power of
a fixed operator, namely $T_{\overline{\eta}} $.

We now come back to the question whether this Toeplitz quantization is distinct from the
coherent state quantization in \cite{csq}.
The definition in equation (24) in \cite{csq} of the coherent state quantization of the `function' $f \in PG_{l,q}$
amounts to the following with our conventions:
$$
A_f := \sum_m w_{l-1-m} \int \!\!\! \int  \mathrm{d} \theta \, | \theta \rangle \,  
\theta^m f(\theta, \overline{\theta}) \, \overline{\theta}{}^m \,
\langle  \overline{\theta} | \, \mathrm{d}   \overline{\theta}.
$$
Here we are using these definitions from \cite{csq} for the coherent states:
\begin{eqnarray*}
| \theta \rangle := \sum_r \dfrac{1}{w_r^{1/2}} \theta^r \otimes | e_r \rangle \in 
\mathcal{B}_H (\theta) \otimes \mathcal{H},
       \\
       \langle  \overline{\theta} | :=  \sum_s \dfrac{1}{w_s^{1/2}} \overline{\theta}{}^s \otimes \langle e_s | \in 
\mathcal{B}_{AH} ( \overline{\theta} ) \otimes \mathcal{H}^\prime.
\end{eqnarray*}
In these formulas  $\mathcal{H}$ is an auxiliary Hilbert space of finite dimension $l$.
We let $\{ | e_n \rangle \}$ be an orthonormal basis of $\mathcal{H}$,
and $\{ \langle e_n | \}$ be its dual orthonormal basis in the dual space $\mathcal{H}^\prime$.
So $A_f$ acts in $\mathcal{H}$, which is to say that  $A :  PG_{l,q} \to \mathcal{L}(\mathcal{H})$.
We now calculate $A_f$ explicitly for $f = \theta^i \overline{\theta}{}^j$ since this was not written out
in \cite{csq}, though certain special cases were shown there.
Suppressing the tensor product notation as in \cite{csq} we obtain
\begin{eqnarray*}
           &&A_{\theta^i \overline{\theta}{}^j} = 
            \sum_m w_{l-1-m} \int \!\!\! \int \mathrm{d} \theta \, | \theta \rangle \,  \theta^m 
            \theta^i \overline{\theta}{}^j
            \overline{\theta}{}^m 
            \, \langle \overline{\theta} | \, \mathrm{d}   \overline{\theta}
\\
&=& 
            \sum_m w_{l-1-m} \int \!\!\! \int \mathrm{d} \theta \left( \sum_r \dfrac{1}{w_r^{1/2}} \theta^r | e_r \rangle \right)
            \theta^{m+i} 
            \overline{\theta}{}^{m+j} 
            \left(  \sum_s \dfrac{1}{w_s^{1/2}} \overline{\theta}{}^s \langle e_s | \right)
           \mathrm{d}   \overline{\theta}
\\
&=& 
            \sum_m w_{l-1-m} \sum_{r s} \dfrac{1}{w_r^{1/2} w_s^{1/2} } \int \!\!\! \int \mathrm{d} \theta  \,
            \theta^{m+i+r} 
            \overline{\theta}{}^{m+j+s} 
            \mathrm{d}   \overline{\theta} \, | e_r \rangle \langle e_s |
\\
&=&
            \sum_m w_{l-1-m} \sum_{r s} \dfrac{1}{w_r^{1/2} w_s^{1/2} } 
            \delta_{l-1,m+i+r} 
            \delta_{l-1,m+j+s} 
         | e_r \rangle \langle e_s |     
\\
&=& 
            \sum_{r s} \dfrac{1}{w_r^{1/2} w_s^{1/2} } 
            \left( \sum_m w_{l-1-m} \chi_l(m)
            \delta_{l-1,m+i+r} 
            \delta_{l-1,m+j+s} \right)
         | e_r \rangle \langle e_s |              
\\
&=& 
            \sum_{r s} \dfrac{1}{w_r^{1/2} w_s^{1/2} } w_{i+r} \chi_l(r) \chi_l(i+r) \delta_{i+r, j+s}
         | e_r \rangle \langle e_s |                        
\\
&=& 
            \sum_{s} \dfrac{1}{w_{j-i+s}^{1/2} w_s^{1/2} } w_{j+s} \chi_l(j-i+s) \chi_l(j+s)
         | e_{j-i+s} \rangle \langle e_s |    .
\end{eqnarray*}
In an equivalent notation (with $e_a = | e_a \rangle$) this reads as
\begin{equation}
\label{equiv-mapsto}
  A_{\theta^i \overline{\theta}{}^j} : e_a \mapsto  \dfrac{w_{j+a} }{w_{j-i+a}^{1/2} w_a^{1/2} } \chi_l(j-i+a) 
  \chi_l(j+a) e_{j-i+a} 
\end{equation}
for all $a \in I_l$.
Note that the left and right versions of the coherent state quantization in \cite{csq}, denoted by
$  A_{\theta^i \overline{\theta}{}^j}^L$ and $  A_{\theta^i \overline{\theta}{}^j}^R$ respectively, give the same expression
modulo some factors of $q$.

We now compare the formula (\ref{equiv-mapsto}) with (\ref{useful}) which we write equivalently as
$$
T_{\theta^i \overline{\theta}{}^j} : \phi_a \mapsto  \dfrac{w_{i+a} }{w_{i-j+a}^{1/2} w_a^{1/2} } \chi_l(i-j+a) \chi_l(i+a) \phi_{i-j+a} 
$$
using the orthonormal basis $\phi_a$ of $\mathcal{B}_H(\theta)$.
This assumes the form of equation (\ref{equiv-mapsto}) provided that we interchange $i$ and $j$, which corresponds
to replacing $\theta^i \overline{\theta}{}^j$ with its conjugate $(\theta^i \overline{\theta}{}^j)^* = \theta^j \overline{\theta}{}^i$.
This motivates the definition of a \textit{linear} map $Z:  PG_{l,q} \to  PG_{l,q}$ defined by $f \mapsto f^*$ for the
elements $f$ in the basis $AW$. 
Please note that $Z$, being the linear extension, is \textit{not} the conjugation.
Also we define a unitary map $U: \mathcal{B}_H(\theta) \to \mathcal{H}$ by $\phi_a \mapsto e_a$ 
and its induced $C^*$-algebra isomorphism
$\tilde{U} : \mathcal{L} (\mathcal{B}_H(\theta) ) \to \mathcal{L} (\mathcal{H})$ given by 
$\tilde{U} : T \mapsto U T U^*$ for all $T \in \mathcal{L} (\mathcal{B}_H(\theta) )$.
So we have proved the following.
\begin{theorem}
The coherent state quantization $A$ and the Toeplitz quantization $T$ are related by
$  A Z  = \tilde{U} T$.
\end{theorem}

\begin{corollary} (Proved in \cite{csq}, Section~7.) 
The coherent state quantization
$
A : PG_{l,q} \to \mathcal{L} (\mathcal{H})
$
is a vector space isomorphism, where we are using the notation established above.
\end{corollary}
\textbf{Proof}: The linear map $Z$ is invertible, so we can write $  A  = \tilde{U} T Z^{-1}$.
This exhibits $A$ as the composition of three vector space isomorphisms.
$\quad \blacksquare$

\vskip 0.2cm
Whether the previous theorem says that these two quantizations are equivalent
will depend on one's definition of equivalence of quantizations. 
That in turn will depend on one's definition of quantization. 
I would rather not give a definition of quantization in general, let alone in the context of this paper.
However, we made some rather arbitrary choices in our definition of a Toeplitz operator.
Another reasonable definition is 
$
     T^\flat_g := P_{\overline{K}} M^L_g,
$
where $M^L_g$ is multiplication on the
left by $g \in PG_{l,q}$ and
$$
P_{\overline{K}} : PG_{l,q}(\theta, \overline{\theta}) \to \mathcal{B}_{AH} (\overline{\theta})
$$ 
is the projection associated to the reproducing kernel $\overline{K}$ of $\mathcal{B}_{AH } (\overline{\theta})$.
(See \cite{part1}.)
Then the Toeplitz quantization $  T^\flat : PG_{l,q} \to \mathcal{L} ( \mathcal{B}_{AH} (\overline{\theta}) )$
has properties corresponding to those of $T$, though some details undergo minor changes.
In particular, as the reader can verify, instead of (\ref{useful}) we obtain
$$
T^\flat_{\theta^i \overline{\theta}{}^j } : \phi_a^* (\overline{\theta}) 
\mapsto \dfrac{w_{j+a} }{w_{j-i+a}^{1/2} w_a^{1/2} } \chi_l(j-i+a) 
  \chi_l(j+a) \phi_{j-i+a}^* (\overline{\theta}).
$$
So in \textit{any} reasonable definition of equivalence of quantizations 
we have: 

\begin{theorem}
The quantizations $A$ and $T^\flat$ are equivalent.
\end{theorem}

In the next result we consider the adjoint operator $(T_g)^*$ of a Toeplitz
operator $T_g \in \mathcal{L} ( \mathcal{B}_H(\theta) )$.
\begin{theorem}
\label{adjoint-theorem}
For all $g \in PG_{l,q}$ we have $(T_g)^* = T_{g^*}$ in $\mathcal{L} ( \mathcal{B}_H(\theta) )$.
\end{theorem}
\textbf{Remark}: While there probably is no such thing as the `right' quantization nor is any definition
of the conjugation in $PG_{l,q}$ the `correct' one, it does turn out that our Toeplitz quantization and our
conjugation are compatible.

\vskip 0.4cm \noindent
\textbf{Proof}:
We first consider the case $g = \eta^i \overline{\eta}{}^j$, an arbitrary element in the basis $AW$.
We use the elements $h_a (\theta) = \theta^a$, which form an orthogonal basis
of $\mathcal{B}_H(\theta)$.
Using equation (\ref{useful}) we calculate matrix elements of $(T_g)^*$ as follows:
\begin{eqnarray*}
\langle h_a , ( T_{\eta^i \overline{\eta}{}^j})^*  h_b \rangle_w &=& 
\langle T_{\eta^i \overline{\eta}{}^j} h_a , h_b \rangle_w
\\
 &=& \left\langle \dfrac{w_{i+a}}{w_{i+a-j} } \chi_l(i+a) \chi_l(i+a-j)  \theta^{i+a-j} , \theta^b \right\rangle_w
\\
 &=& \dfrac{w_{i+a}}{w_{i+a-j} } \chi_l(i+a) \chi_l(i+a-j)  \left\langle  \theta^{i+a-j} , \theta^b \right\rangle_w
\\
 &=& \dfrac{w_{i+a}}{w_{i+a-j} } \chi_l(i+a) \chi_l(i+a-j) \, w_{i+a-j} \, \delta_{i+a-j , b}
\\
 &=& w_{i+a} \, \chi_l(i+a) \chi_l(i+a-j) \, \delta_{i+a-j , b}.
\end{eqnarray*}
Next we calculate matrix entries of $ T_{g^*}$, again using equation (\ref{useful}):
\begin{eqnarray*}
\langle h_a ,  T_{\eta^j \overline{\eta}{}^i}  h_b \rangle_w &=& 
\left\langle \theta^a,  \dfrac{w_{j+b}}{w_{j+b-i} } \chi_l(j+b) \chi_l(j+b-i) \theta^{j+b-i} \right\rangle_w
\\
&=& \dfrac{w_{j+b}}{w_{j+b-i} } \chi_l(j+b) \chi_l(j+b-i) \left\langle \theta^a, \theta^{j+b-i} \right\rangle_w
\\
&=& \dfrac{w_{j+b}}{w_{j+b-i} } \chi_l(j+b) \chi_l(j+b-i) \delta_{a,j+b-i} w_{j+b-i}
\\
&=& w_{j+b} \chi_l(j+b) \chi_l(j+b-i) \delta_{a,j+b-i}.
\end{eqnarray*}
Because of the Kronecker deltas, the only non-zero values for these two matrix elements
occur for $i+a = j + b$.
But in that case we have $\chi_l(i+a-j) = \chi_l (b) = 1 $
and $ \chi_l(j+b-i) = \chi_l (a) =1$.
And the other factors in these two expressions for the matrix elements are also
equal. So we conclude that
$$
       ( T_{\eta^i \overline{\eta}{}^j})^* = T_{\eta^j \overline{\eta}{}^i} = T_{ ( \eta^i \overline{\eta}{}^j )^* } \, .
$$
This proves the theorem for the special case when $g = \eta^i \overline{\eta}{}^j$.
The proof for a general element $g$ follows immediately by expanding $g$ in the
basis $AW$.
$\quad \blacksquare$

\vskip 0.4cm 
For the corollary of this result we first recall a standard definition.
\begin{definition}
An element $r$ in a $*$-algebra is {\em self-adjoint} or  {\em real}  if $r^* = r$.
\end{definition}
\begin{corollary}
The Toeplitz operator  $T_g$ is self-adjoint  if and only if $g$ is self-adjoint.
\end{corollary}
\textbf{Proof}: If $g^* = g$, then by the previous theorem we have $(T_g)^* = T_{g^*} = T_g$.
Conversely, $T_g$ self-adjoint implies $T_g = (T_g)^* = T_{g^*}$, again by the previous theorem.
Since $T$ is injective we get $g = g^*$ as desired. 
$\quad \blacksquare$

\vskip 0.4cm 
While $T$ is not an algebra morphism, it does have some nice properties with respect to
the products.
The following is a typical property of Toeplitz operators.
The proof is essentially the same as in the context of reproducing kernel Hilbert spaces of functions.
\begin{theorem}
Suppose that $g_1, g_2 \in \mathcal{B}_H(\eta)$. Then 
$
      T_{g_1} T_{g_2}  =   T_{g_1 g_2}   = T_{g_2} T_{g_1}  .
$
For $h_1, h_2 \in \mathcal{B}_{AH}(\overline{\eta})$ one has 
$
           T_{h_1} T_{h_2} = T_{h_1 h_2} =  T_{h_2} T_{h_1}  .
$
So $T$ restricted to either $\mathcal{B}_H(\eta)$ or $\mathcal{B}_{AH}(\overline{\eta})$ is an algebra homomorphism.
\end{theorem}
\textbf{Proof}: 
Using $T_g =P_K M_g$ and $M_{g_1} M_{g_2} = M_{g _2 g_1}$ we have that
$$
 T_{g_1} T_{g_2} = P_K M_{g_1} P_K M_{g_2}  = P_K M_{g_1} M_{g_2} =  P_K M_{g _2 g_1} =
 T_{g _2 g_1}.
$$
Interchanging $g_1$ and $g_2$ gives $T_{g_2} T_{g_1} = T_{g _1 g_2}$.
But $g _2 g_1 = g _1 g_2$ giving $ T_{g_1} T_{g_2} =   T_{g_1 g_2} = T_{g_2} T_{g_1} $
The adjoint of this is $ T_{g_2^*} T_{g_1^*} =   T_{(g_1 g_2)^*} = T_{g_1^*} T_{g_2^*}$.
Now 
$(g_1 g_2)^* = g_1^* g_2^*$ (even though 
$PG_{l,q}$ need not be a $*$-algebra) as the reader can show. 
But we can write any $h \in \mathcal{B}_{AH}(\overline{\eta})$ as $h = g^*$
with $g \in \mathcal{B}_{H}(\eta)$ and so $ T_{h_1} T_{h_2} = T_{h_1 h_2} =  T_{h_2} T_{h_1}$ then follows.
$\quad \blacksquare$

\vskip 0.4cm 
We now consider the commutation relation between $M_\theta$ and $\partial_w$.
First
\begin{equation}
\label{e-values}
     (M_\theta \partial_w) \theta^a = M_\theta \left( \dfrac{w_a}{w_{a-1}}   \chi_l(a-1)  \right)  \theta^{a-1}
     = \dfrac{w_a}{w_{a-1}} \chi_l(a-1) \, \theta^a
\end{equation}
holds for $a \in I_l$.
In the other order we get
$$
    (\partial_w M_\theta) \theta^a = \partial_w \left( \chi_l(a+1) \, \theta^{a+1} \right)
    = \dfrac{w_{a+1}}{w_{a}} \chi_l(a+1) \, \theta^a.
$$
In general, this does not lead to a simple formula for the commutator unless we suppose
that the weights $w_n$ satisfy some other relations.
Nonetheless, by (\ref{useful}) 
we do have
$$
   T_{\eta \overline{\eta}} \theta^a = \dfrac{w_{a+1}}{w_{a}} \, \chi_l(a+1) \theta^a.
$$
And so $T_{\eta \overline{\eta}} =  \partial_w M_\theta = T_{\overline{\eta}} T_{\eta}$.
We can also write this, strangely enough, as 
$q T_{\overline{\eta} \eta} = T_{q \overline{\eta} \eta} = T_{\eta \overline{\eta}}  = T_{\overline{\eta}} T_{\eta}$.
And again, in general, this has no simple relation with $T_{\eta}  T_{\overline{\eta}} = M_\theta \partial_w $
as we noted above.

We now note that by Theorem \ref{adjoint-theorem} these two operators
$ \partial_w = T_{\overline{\eta}} = T_{\eta^*} $
and $M_\theta =  T_{\eta}$ are adjoints in  $\mathcal{L} ( \mathcal{B}_H(\theta) )$
with respect to the sesquilinear form.
The operators $M_\theta $ and $M_{\overline{\theta}}$ (both defined as multiplication on the right acting on
$PG_{l,q}$) satisfy this \textit{$q$-commutation relation} in $PG_{l,q} (\theta, \overline{\theta})$:
$$
  M_{\overline{\theta} } M_{\theta} - q M_{\theta} M_{\overline{\theta} } = 0.
$$
However, the projection $P_K$ from $PG_{l,q}$ to $\mathcal{B}_H (\theta)$ defined
by the reproducing kernel is not an algebra homomorphism and so this particular relation
is not necessarily preserved.
Of course, this is to be expected since the parameters $w_a$ in the reproducing kernel in general
have nothing whatsoever to do with the parameter $q$.
(But in \cite{csq} these parameters are related. This can be considered an
advantage of the approach of those authors.)

Operators satisfying the condition $T T^* = q T^* T$ were introduced in \cite{ota}.
They are called \textit{$q$-normal operators}
and  have been studied in \cite{kula-ricard} and \cite{otafr}.
Now $M_\theta $ and $M_{\overline{\theta}}$ are not necessarily adjoint operators
in $PG_{l,q}$. 
But $T_g = \iota^* M_g \iota$, where the inclusion 
$\iota : \mathcal{B}_H \to PG_{l,q}$ is isometric and $\iota^* = P_K$, where we are considering
$$
     P_K : PG_{l,q} \to \mathcal{B}_H.
$$
One says that $T_g= \iota^* M_g \iota$ is the \textit{compression} of $M_g$ to $\mathcal{B}_H$.
We feel that the following definitions have been given some inspiration by this discussion and
the material in \cite{conway}.
\begin{definition}
Let $q \in \mathbb{C}$.
Let $A, B \in \mathcal{L}(X)$ for some Hilbert space $X$.
We say that $A$ and $B$ {\rm $q$-commute up to compression} if there exists a complex inner product space
$X^\prime$ containing $X$ and 
 $A^\prime$, $B^\prime \in \mathcal{L}(X^\prime)$ whose compressions to $X$
 are $A$ and $B$, respectively, and  such that
$A^\prime$, $B^\prime$ are \textit{$q$-commuting operators} in $X^\prime$, that is 
$A^\prime B^\prime = q B^\prime A^\prime $.
Here the {\rm compression} of $A^\prime$ to $X$ is defined to be $\iota^* A^\prime \iota$, where
the inclusion $\iota : X \to X^\prime$ is  isometric.
We say that $A$ is a {\rm sub-$q$-normal operator} if $A$ and $A^*$ $q$-commute up to compression.
\end{definition}

Note that neither the space $X^\prime$ nor the pair of operators $A^\prime$, $B^\prime $ will be unique.
Given this terminology we have that the pair of operators $\partial_w, M_\theta $ (in that order!)
acting on $\mathcal{B}_H (\theta)$ $q$-commute up to compression.
The point here is that these types of operators possibly could be new and therefore
of interest to researchers in operator theory
(cp. \cite{conway}).
However, we do not elaborate on this for now.

Taking the point of view of physics, we also define a \textit{creation operator}
by $A^\dagger_w :=  M_\theta = T_\eta$, which we now know is the adjoint of
the \textit{annihilation operator} $A_w := \partial_w = T_{\overline{\eta}}$.
Though this terminology comes from physics, the mathematical fingerprint of
an annihilation operator is that it lowers the degree (of homogeneous elements) by $1$
while a creation operator raises the degree by $1$.
For this reason they are often called the lowering and raising operators, respectively.
Thus in this Toeplitz quantization, it is unambiguous that $\eta$ corresponds to the
creation operator and that $\overline{\eta}$ corresponds to the annihilation operator.
So the quantization scheme breaks the symmetry in the roles of $\eta$ and $\overline{\eta}$
in the pre-quantization space $PG_{l,q } (\eta,\overline{\eta})$.
But there is another, quite natural Toeplitz quantization scheme available in this set-up.
This is the linear map 
$$
PG_{l,q}(\eta, \overline{\eta}) \to \mathcal{L} ( \mathcal{B}_{AH}(\overline{\theta} ) )
$$
defined by multiplying elements in the anti-Segal-Bargmann space $\mathcal{B}_{AH}(\overline{\theta} )$
 on the right by a fixed element in the symbol space $
PG_{l,q}(\eta, \overline{\eta})$ and then projecting back into the anti-Segal-Bargmann space
by using its reproducing kernel `function'.
This gives a theory that is isomorphic (or anti-isomorphic, depending on your definitions) to what
we have presented.
Now by applying the above `fingerprint' test we see that
this new Toeplitz quantization breaks the symmetry in the roles of
$\eta$ and $\overline{\eta}$ in $PG_{l,q } (\eta,\overline{\eta})$
by sending $\eta$ to an annihilation operator while sending $\overline{\eta}$ to a creation operator.
And this is the reverse of what happens in our Toeplitz quantization!
Actually, naming the two subspaces $\mathcal{B}_{H}(\theta )$ and
$\mathcal{B}_{AH}(\overline{\theta} )$ as holomorphic and anti-holomorphic, respectively, is totally
arbitrary and is most likely due to a desire to maximize creature comfort more than anything else. 
In other words, $\overline{\theta}$ is just as good a complex variable as $\theta$ and so the names
of these two subspaces can be interchanged with no damage to mathematical ideas.

Toeplitz quantization in the original context of Segal-Bargmann analysis 
is related to the anti-Wick quantization. (See \cite{hall}.)
To see what is happening in the current set-up, we take elements $\phi \in \mathcal{B}_{H} (\theta)$
and $\psi \in \mathcal{B}_{AH} (\overline{\theta})$.
Using $T_g = P_K M_g$ and $M_{g_1 g_2} = M_{g_2} M_{g_1}$, we get another multiplicative relation:
$$
T_{\phi \psi} = P_K M_{\phi \psi} = P_K M_\psi M_\phi =  P_K M_\psi P_K M_\phi = T_\psi T_\phi.
$$
The second $P_K$ in the fourth expression acts as the identity since the last factor $M_\phi$ leaves
$\mathcal{B}_H(\theta)$ invariant.
In particular, on a basis element in $AW$ we get
\begin{equation}
\label{in-anti-wick-order}
  T_{\eta^i \overline{\eta}{}^j} = T_{ \overline{\eta}{}^j}   T_{\eta^i }  = (T_{ \overline{\eta} })^j  (T_{\eta})^i
  = (A_w)^j (A_w^\dagger)^i,
\end{equation}
which is in anti-Wick order.
So for an arbitrary symbol $g$ we use linearity to write $T_g$ as a linear combination of terms
in anti-Wick order.
We can change this Toeplitz quantization by using the left multiplication operator instead of the right.
Also we can use the anti-Segal-Bargmann space as noted above, and in this case we have two choices for the multiplication
operator as well.
In all of these other Toeplitz type quantizations, we get an anti-Wick expression of the form in (\ref{in-anti-wick-order}).
Moreover, we have this interesting consequence.
\begin{theorem}
The set $\{ (A_w)^j (A_w^\dagger)^i \, | \, i,j \in I_l \}$ of anti-Wick ordered elements
is a basis of $\mathcal{L}(\mathcal{B}_H (\theta))$.
\end{theorem}
\textbf{Proof}: By equation (\ref{in-anti-wick-order}) this set of linear maps is the image under the
vector space isomorphism $T$ of the basis $AW$.
$\quad \blacksquare$

\vskip 0.4cm \noindent
\textbf{Remark}: Since we lack simple commutation relations for the creation and annihilation
operators, we do not make any statement for now about whether the set $\{ (A_w^\dagger)^i (A_w)^j  \, | \, i,j \in I_l \}$
of Wick ordered elements is also a basis.
This is a most curious situation which merits further consideration.

\vskip 0.4cm \noindent
The \textit{number operator} $N_w$ in this context is defined to be
$$
N_w := A_w^\dagger A_w = M_\theta \partial_w = T_\eta T_{\overline{\eta}} = T_\eta T_\eta ^*.
$$
Clearly, $N_w \ge 0$.
Of course,  $A^\dagger_w , A_w, N_w$ are operators in the Hilbert space $\mathcal{B}_H(\theta)$.
We have shown in (\ref{e-values}) that the basis $\theta^a$ with $a \in I_l$ diagonalizes the operator $N_w$
and that its spectrum is $\{ w_a / w_{a-1}  ~|~ a \in I_l  \}$,
where $w_0 / w_{-1} \equiv 0$.
$N_w$ serves as a Hamiltonian operator in this theory.
One can think of $N_w$ as a deformed harmonic oscillator Hamiltonian, possibly up to
an additive constant.
It defines a \textit{Dirichlet form} 
$ \langle f, N_w f \rangle_w = || A_w f ||_w^2$
for all $f \in \mathcal{B}_H$.
We can define $[a]_w := w_a / w_{a-1}$, the \textit{w-deformation} of the integer $a \in I_l$.
A $w$-deformation of the factorial function can also be introduced.
Note that in this formulation, commutation relations for the operators $A^\dagger_w , A_w, N_w$
do not necessarily play a role. 
The material in this paragraph gives some notational concordance with other papers.

\section{Concluding Remarks}

We consider the theory of Toeplitz operators developed in this paper to be just the beginning of a longer story
that should eventually include results for infinite dimensional quantum spaces with a reproducing kernel.
This will allow one to consider other properties that are not relevant in finite dimensions, such as
whether the Toeplitz operator for a given symbol is bounded, compact, in a Schatten class and so forth.
 
 The `functions' in this paper are not functions, but elements in an algebra.
 We do think of these elements as arising from a functional calculus, of course. 
 Here the algebras of `functions' is $PG_{q,l}$, which is not
 commutative (except in the case $q=1$).
 So the Toeplitz quantization $g \mapsto T_g$ for $g \in PG_{q,l}$  is analogous to \textit{second quantization} in physics,
 where one quantizes a theory that is itself a quantum (that is, non-commutative) theory to begin with.
We also expect that other second quantizations of interest in mathematics and physics can be analyzed along
the lines indicated in this paper.

We have only considered in this paper the case of one pair of complex paragrassmann variables.
 In \cite{csq} the generalization to a finite family of such pairs is given.
It is plausible to conjecture that their holomorphic subspaces have reproducing kernels and associated
Toeplitz operators in that case, too.
Also we have a reproducing kernel for $PG_{q,l}  (\theta, \overline{\theta})$ (see \cite{part1}),
and this space can be embedded into any space with a finite family of complex paragrassmann variables.
So there could very well be a theory of Toeplitz operators acting on $PG_{q,l} (\theta, \overline{\theta})$
for such embeddings.

The nilpotency conditions gave us a finite dimensional space.
It seems reasonable that Toeplitz operators could also be defined on the
\textit{quantum plane} defined by 
$\mathbb{C}  \{ \theta, \overline{\theta} \} / \langle \theta \overline{\theta} - q \overline{\theta} \theta \rangle$.
This is a well known and studied object in non-commutative geometry, but this would be way of viewing it
from the perspective of analysis.
The motivation for this paper as well as for Part I (see \cite{part1}) is to introduce ideas from analysis into the study of
non-commutative spaces.
We expect that the ideas of reproducing kernels and their associated Toeplitz operators,
as well as other aspects of analysis, will find more applications in non-commutative geometry.

\vskip 0.4cm \noindent
\textbf{\Large Acknowledgments}
\vskip 0.4cm \noindent
As with Part I, this paper was also inspired by a talk based on \cite{csq} given by Jean-Pierre Gazeau during my sabbatical stay
at the Laboratoire APC,  Universit\'e Paris Diderot (Paris 7) in the spring of 2011. 
I am extremely grateful to Jean-Pierre for both his inspiration and his hospitality.
Merci beaucoup, Jean-Pierre!
And again, as in Part I, I wish to thank Rodrigo Fresneda for most useful conversations
during my stay at the UFABC in S\~ao Paulo, Brazil in April, 2012 and for being my most gracious host there.
Muito obrigado, Rodrigo!
Finally, I most warmly thank Anna Kula for making me aware of her work on $q$-normal operators
and other related work as well as for
 providing me with a copy of reference \cite{kula-ricard}.
Dzi\c ekuj\c e bardzo, Anna!

\end{document}